\documentclass[prl,twocolumn,nofootinbib,superscriptaddress,10pt]{revtex4-1}

\usepackage{graphicx}
\usepackage{graphics}
\usepackage{amsmath,amssymb}
\usepackage[usenames]{color}

\def\mysection#1{{{\bf #1}.~}}

\newcommand{\eps}{\epsilon}

\newcommand{\lqcd}{\Lambda_{\rm QCD}}
\newcommand{\mpl}{M_{\rm Pl}}

\newcommand{\be}{\begin{equation}}
\newcommand{\ee}{\end{equation}}
\newcommand{\bea}{\begin{eqnarray}}
\newcommand{\eea}{\end{eqnarray}}
\newcommand{\beq}{\begin{equation}}
\newcommand{\eeq}{\end{equation}}
\newcommand{\beqa}{\begin{eqnarray}}
\newcommand{\eeqa}{\end{eqnarray}}

\renewcommand{\eps}{{\epsilon}}

\definecolor{BrickRed}{cmyk}{0,0.89,0.94,0.28}
\definecolor{MidnightBlue}{cmyk}{0.98,0.13,0,0.43}
\definecolor{DarkGreen}{rgb}{0.100806,0.495968,0.209979}
\definecolor{orange}{rgb}{0.587167,0.354498,0.146197}

\begin{document}

\title{~~\\ Why do we observe a weak force? \\
The hierarchy problem in the multiverse}

\author{Oram Gedalia} \address{Department of Particle Physics \&
Astrophysics, Weizmann Institute of Science, Rehovot 76100, Israel}

\author{Alejandro Jenkins} \address{Physics Department, Florida State
University, Tallahassee, Florida 32306-4350, USA}

\author{Gilad Perez} \address{Department of Particle Physics \&
Astrophysics, Weizmann Institute of Science, Rehovot 76100, Israel}

\begin{abstract}
Unless the scale of electroweak symmetry breaking is stabilized
dynamically, most of the universes in a multiverse theory will
lack an observable weak nuclear interaction. Such ``weakless
universes'' could support intelligent life based on organic
chemistry, as long as other parameters are properly adjusted.
By taking into account the seemingly-unrelated flavor dynamics
that address the hierarchy of quark masses and mixings, we show
that such weakless (but hospitable) universes can be far more
common than universes like ours.  The gauge hierarchy problem
therefore calls for a dynamical (rather than anthropic)
solution.
\end{abstract}

\maketitle

\mysection{Bayesian statistics for multiverse} The common
wisdom is that, due to the vast landscape of configurations
that are local energy minima, string theory does not uniquely
predict the spectrum of particles and interactions as observed
in our Universe~\cite{douglaskachru}.  Eternal inflation might
then generate an enormous number of causally-disconnected
``pocket universes,'' each with its own laws of
physics~\cite{eternal}. It is unlikely that we shall ever study
directly the physics of universes other than our own, but a
limited Bayesian analysis is still possible in such a
scenario~\cite{carter}: Let $\{\alpha_i \}$ be the set of
parameters of the Standard Model (SM), and let
$\{\alpha_i^\oplus \}$ represent the values that we have
deduced from experiments in our own Universe. The theory
describing the multiverse (whether it be string theory or
something else) in principle gives a probability density
function (PDF), $p(\{\alpha_i \})$, corresponding to the
fraction of universes with those parameters. By Bayes's
theorem, the probability of measuring a set of parameters
$\{\alpha_i \}$ is
\bea
p_{\rm measured} ( \{ \alpha_i \} ) &\equiv& p ( \{\alpha_i \}
| {\rm observer} ) \nonumber \\ &\propto& p ( {\rm observer} |
\{ \alpha_i \} ) \cdot p( \{\alpha_i \})~.
\label{eq:anthropics}
\eea
The conditional PDF for an observer existing, given a set of
values for the SM parameters, $ p ( {\rm observer} | \{
\alpha_i \} )$, is the {\it anthropic} factor in this analysis.
All that we may say with definiteness is that, if
\be
r \equiv \frac{p_{\rm measured}\{\alpha_i^\oplus \}}{p_{\rm
measured}\{\alpha_i^\ast \}} \ll 1~, \label{eq:r-parameter}
\ee
where $\{ \alpha_i^\ast \}$ is a set of parameters well outside
the tolerances of the observed $\{\alpha_i^\oplus \}$, then the
theory makes our own ``way of life'' an unexplained rare event.

\medskip

\mysection{Hierarchies} In the SM, $M_{\rm weak}$ is given by
\be
M_{\rm weak} = \frac{1}{2} g v\,, \label{eq:higgsVEV}
\ee
where $g$ is the weak coupling constant and $v$ is the vacuum
expectation value (VEV) of the Higgs field. Quantum corrections
to (mass)$^2$ of the Higgs depend quadratically on the cutoff
scale, which one would expect to be of the order of $\mpl$ (the
Planck mass), at which quantum gravitational effects become
dominant. As Wilson first stressed~\cite{gaugehierarchy}, this
implies that the observed hierarchy
\be
\frac{M_{\rm weak}}{\mpl} \sim 10^{-16}
\label{eq:weak-hierarchy}
\ee
requires a fine-tuning of the bare (mass)$^2$ of the Higgs
field, to a part in $10^{32}$. A vast amount of work has gone
into proposing theoretical models, such as
technicolor~\cite{TC} and supersymmetry~\cite{SUSY}, that
invoke still undetected interactions in order to obtain the
``gauge hierarchy'' of Eq.~(\ref{eq:weak-hierarchy}) without
fine-tuning.

An even greater hierarchy appears in cosmology: the ``dark
energy'' scale $M_{\rm DE}$, deduced from the acceleration of
the rate of expansion of the Universe~\cite{darkenergy}, is
\be
\frac{M_{\rm DE}}{\mpl} \sim 10^{-31}~.
\label{eq:DE-hierarchy}
\ee
Weinberg argued in~\cite{weinberg1} that the smallness of
$M_{\rm DE}/\mpl$ might have an anthropic explanation: only
very small values of $M_{\rm DE}$ are observable, since
otherwise the dark energy would prevent structure formation in
the Universe, and could therefore be incompatible with the
presence of any conscious observer.  In other words, in
Eq.~\eqref{eq:anthropics}, $p ( {\rm observer} | \{ \alpha_i \}
)$ is zero unless the absolute value of $M_{\rm DE}$ is very
small compared to $\mpl$.  If very small values $M_{\rm
DE}/\mpl$ are at all possible (i.e, if $p( \{\alpha_i \} )$ is
non-zero for such values), it might not require deliberate
fine-tuning to explain why we observe a dark-energy hierarchy.

Does the gauge hierarchy of Eq.~(\ref{eq:weak-hierarchy}) also
admit an anthropic explanation? If all other SM parameters are
held fixed, then $v$ could not be more than about 5 times
greater without destabilizing atoms~\cite{VEV-selection}, and
such conditions might well be incompatible with intelligent
observers. This argument, however, relies on scanning $v$ while
fixing other SM parameters at their observed values. Whether
this assumption is justified, and what the consequences of
relaxing it are, will be our focus here.

In the SM, a fermion mass $m_f$ depends on both the Higgs VEV
$v$ and the Yukawa coupling $y_f$:
\be
m_f = \frac{1}{\sqrt 2} y_f v\,.
\label{eq:fermions}
\ee
Fermions in the SM are organized into generations
(``flavors''), identical except for their Yukawa couplings. The
SM flavor sector is described by 13 parameters, plagued by
hierarchies that constitute the so-called ``flavor
puzzle.''\footnote{This is usually called a ``puzzle," rather
than a ``problem,'' because the flavor parameters are not
subject to large additive quantum corrections.}  For instance,
one would generically expect $C^{\rm SM}$, the ``Jarlskog
determinant'' that characterizes the amount of charge-parity
violation in the SM, to be $\sim 0.1$, whereas in reality
$C^{\rm SM} \sim 10^{-22}$~\cite{Jarlskog}  (far too small to
explain our Universe's matter-antimatter asymmetry).

Except for rare, high-energy processes, most of the physical
phenomena in our Universe depend on the masses of only the
three lightest fermions: the electron $e$, and the $u$ and $d$
quarks. One might therefore expect that only these masses would
play an important role in anthropic constraints on possible
laws of physics. Thus, the flavor puzzle does not seem to admit
an anthropic solution, calling instead for a {\it dynamical}
solution to account for the observed hierarchies.  Remarkably,
we shall see that by incorporating such flavor dynamics
---which has no known theoretical connection to the weak
scale--- both $v$ and the $y_f$'s are likely to scan over the
multiverse, and that stable atoms become possible (even
favored!) with $v \gg v^\oplus$.

\medskip

\mysection{Weakless universe} Before analyzing the implications
of flavor dynamics in the multiverse, let us briefly summarize
the argument of~\cite{weakless}, as to how to build a
hospitable universe without weak interactions. The SM
parameters of this universe are indicated by a
superscript~$\ast$. Let
\be
v^\ast \simeq \mpl
\label{eq:naturalVEV}
\ee
(i.e., the Higgs VEV takes its most natural value). The gauge
groups and low-energy gauge couplings are the same as in our
Universe and $\lqcd^\ast = \lqcd^\oplus$ (where $\lqcd$ is the
energy scale below which the strong nuclear interaction is
non-perturbative). The Yukawa couplings are:
\be
y_f^\ast = \frac{y_f^\oplus v^\oplus}{v^\ast}, ~~\hbox{for}~~ f
= e, u, d
\label{eq:lightfermions}
\ee
(i.e., $m^\ast_f = m_f^\oplus$, for $f = e, u, d$).

As long as
\be \label{eq:smass}
m_s \gtrsim \frac{m_u + m_d}{2} + 5 ~{\rm MeV} \,,
\ee
the $s$ quark will not participate in forming stable nuclei
(which otherwise could have no more than a couple of units of
electric charge, leading to incompatibility with organic
chemistry)~\cite{quarkmasses}.  For simplicity, we just require
that
\be
m_f^\ast \gtrsim \lqcd^\ast
\ee
for all flavors other than $e,u,$ and $d$.

In the weakless universe, the usual fusing of four protons to
form helium would be impossible, because it requires that two
of the protons convert into neutrons, via the weak force. But
other pathways could exist for nucleosynthesis: A small
adjustment to the parameter that characterizes the
matter-antimatter asymmetry of the Universe, $\eta_b^\ast
\simeq 10^{-2} \eta_b^\oplus \,,$ (where $\eta_b^\oplus \sim
10^{-10}$ is the cosmological baryon-to-photon ratio)  is
enough to ensure that the Big Bang nucleosynthesis would leave
behind a substantial amount of deuterium nuclei. Stars could
then shine by fusing a proton and a deuterium nucleus to make a
helium-3 nucleus~\cite{weakless}. Such a universe could still
produce heavy elements up to iron, stars might be able to shine
for several billion years at an order-one fraction of the
brightness of stars in our own Universe, and type Ia supernova
explosions could still disperse the heavy elements into the
interstellar medium.

In terms of the Bayesian analysis of Eq.~\eqref{eq:anthropics},
it therefore seems that
\be
p ( {\rm observer} | \{ \alpha_i^\oplus \}) \simeq p ( {\rm
observer} | \{ \alpha_i^\ast \})~.
\ee

\mysection{Naturalness} The dependence of $v$ on quantum
corrections to the (mass)$^2$ parameter for the Higgs field
suggests that large values of it are more likely. Since the
sign of $v$ is not physical, the most conservative assumption
is
\be
p(v) \sim \frac{v^2}{\mpl^2} ~~\hbox{for}~~ 0 < v < \mpl~.
\label{eq:naturalness}
\ee
Another way of stating the gauge hierarchy problem is that
Eq.~\eqref{eq:naturalness} suppresses the likelihood of finding
a universe with a Higgs VEV as small as $v^\oplus$, by a factor
of $10^{-32}$. It is unclear at this point how $v$ might be
distributed in the landscape of string
theory~\cite{landscape-v2, Hall&Nomura}, but
Eq.~\eqref{eq:naturalness} agrees with the expectation from the
toy landscape considered in~\cite{friendly-landscape}, if no
symmetry requires the mean value of $v$ to vanish.  In any
case, we shall argue that unless the PDF for $v$ is
significantly less biased toward large values than
Eq.~\eqref{eq:naturalness}, then a multiverse theory could fail
the Bayesian test of Eq.~\eqref{eq:r-parameter}.

\medskip

\mysection{Flavor dynamics} During the past 30 years or so,
theoretical physicists have proposed various mechanisms that
might explain the flavor hierarchy
puzzle~\cite{froggatt-nielsen, strong-dynamics,
split-fermions}. The resulting wisdom regarding the generation
of flavor hierarchies can be summarized by the following
relation between the fundamental flavor parameters and the
effective Yukawa couplings:
\be
y \propto \eps^Q~,
\label{eq:fn_yukawa}
\ee
where $\eps$ is some small universal parameter and $Q$ is a
flavor-dependent charge. (In these models, the values of $y$
are usually related to the VEV of a scalar field.) Consider a
general power-law distribution for the charge $Q$ on the
multiverse,
\be
p(Q) \propto Q^n~,
\label{eq:q_pdf}
\ee
and let $\eps$ have an arbitrary distribution $p(\eps)$. Then
the PDF for $y$ is
\bea
p(y) &\propto& \int d \eps \, dQ \, p(\eps) Q^n \delta( y -
\eps^Q) \nonumber \\ && ~~~   =  \int d \eps \, dQ \, p(\eps) Q^n
\frac{ \delta( Q - \ln y / \ln \eps) } {y \ln
\eps} \nonumber \\
&& ~~~  = \left( \frac{\ln^n y }{y} \right) \times \int d \eps \,
\frac{p(\eps)}{\ln^{n+1} \eps} \propto \frac{\ln^n y}{y}~.
\label{eq:yuk_pdf}
\eea

Let us evaluate the $r$ of Eq.~\eqref{eq:r-parameter} as the
ratio of hospitable universes with $v = v^\oplus$ (i.e., weak
universes, like our own) and $v = v^\ast \sim \mpl$ (i.e.,
weakless universes), with all other parameters integrated over.
This $r$ has two factors: the first is the ratio of
probabilities for the corresponding values of $v$, which by
Eq.~\eqref{eq:naturalness} is simply $(v^\oplus/\mpl)^2$.  The
second is the ratio of the probabilities for the Yukawa
couplings that produce the corresponding quark masses (see
Eq.~\eqref{eq:fermions}).

By Eq.~\eqref{eq:yuk_pdf}, each light quark (i.e., each quark
with a mass below $\lqcd$) contributes to $r$ a factor of
\be \label{eq:lightratio}
\frac{\int_0^{\lqcd / v^\oplus} dy \, p(y)}{\int_0^{\lqcd /
\mpl} dy \, p(y)} = \frac{\left[ \ln ^{n+1} y \right]^{\lqcd /
v^\oplus}_0} {\left[ \ln ^{n+1} y \right]^{\lqcd / \mpl}_0}~,
\ee
which is equal to 1 for $n \geq -1$, or to
\be
\left[ \frac{\ln \left( \frac{\lqcd}{\mpl} \right)}{\ln
\left( \frac{\lqcd}{v^\oplus} \right)} \right]^{-n-1}
\approx ~ 6.7^{-n-1}
\ee
for $n < -1$.  On the other hand, for a single heavy quark, the
corresponding factor is less than or equal to 1, for any $n\,$.

Thus, the full ratio $r$ of weak to weakless universes with two
light quarks ---which are necessary to get organic
chemistry~\cite{quarkmasses}--- and any number of heavy quarks,
can be bounded by
\be \label{eq:finalratio}
r< \left(\frac{v^\oplus}{\mpl}\right)^2 \times \left\{
\begin{array}{ll} \left[ \frac{\ln \left(
\frac{\lqcd}{\mpl} \right)}{\ln \left( \frac{\lqcd}{v^\oplus}
\right)} \right]^{-2n-2} & n<-1 \,, \\ 1 & n \geq -1 \,.
\end{array} \right.
\ee
We therefore conclude that $r \ll 1$ for any $n \gtrsim -21$.
Unless $\epsilon$ were made unnaturally small (which would be a
hierarchy problem of its own), a model with $n < -21$ would
strongly favor large Yukawa couplings and would conflict with
the observed distribution of fermion masses.\footnote{The
anthropic requirement that the electron be
light~\cite{Hall&Nomura,Zakopane} can be accommodated in this
analysis, without significantly altering the conclusions.}

\begin{figure} [t]
\begin{center}
\includegraphics[width=0.45\textwidth]{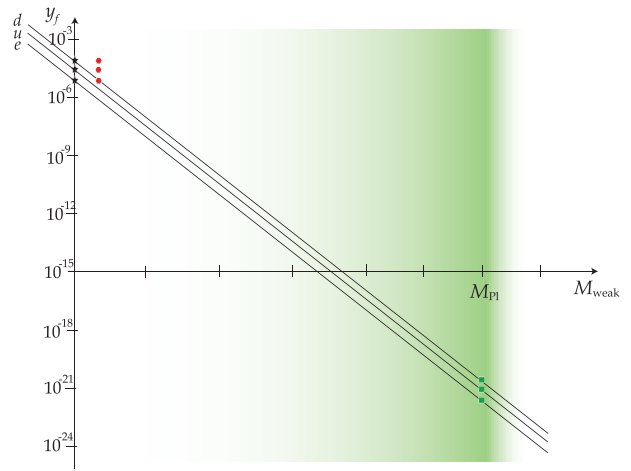}
\end{center}
\caption{\footnotesize Log-log plot of the Yukawa couplings $y_f$ as
functions of $M_{\rm weak}$.  Black stars correspond to our Universe,
red hexagons to a spectrum incompatible with stable atoms~\cite{VEV-selection}
and green squares to the weakless universe described by
Eqs.~\eqref{eq:naturalVEV} and~\eqref{eq:lightfermions}. The green shading
roughly indicates the expected probability density in the multiverse,
from Eqs.~\eqref{eq:naturalness} and~\eqref{eq:scaleinvariance}.}
\label{fig:yukawas}
\end{figure}

For $n=0$, Eq. \eqref{eq:yuk_pdf} becomes
\be
p(y) \propto \frac{1}{y}~,
\label{eq:scaleinvariance}
\ee
which corresponds to a {\it scale-invariant} distribution, in
which $y$ is uniformly distributed on a logarithmic
scale.\footnote{It is easy to show that ---other than
$\delta(x)$--- $1/x$ is the only scale-invariant PDF, but it is
not normalizable, and needs to be cut off at some minimum and
maximum values.} Approximate scale-invariance (between cutoffs)
is a common feature of complex dynamics and leads, for
instance to ``Benford's law,'' the well-known statistical
observation that the first digits of data from a surprisingly
wide variety of sources are logarithmically distributed, with
the digit 1 being six times more likely than the digit
9~\cite{benford}. Approximate scale-invariance is therefore a
plausible expectation for something like the distribution of
Yukawa couplings in the multiverse, independently of specific
flavor models. Figure~\ref{fig:yukawas} illustrates why, if the
Yukawa couplings have a scale-invariant distribution,
Eq.~\eqref{eq:naturalness} would imply that we would be far
more likely to find ourselves living in a weakless universe,
rather than in one with a measurable weak nuclear interaction.

The authors of~\cite{Donoghue:2009me} consider a PDF for $y$ of
the form of Eq.~\eqref{eq:scaleinvariance}, which requires
cutoffs. A lower cutoff can disallow the $y_f^\ast \sim
10^{-21}$ of the weakless universe. There is no known dynamical
reason, however, why such values would be forbidden in the
string landscape.\footnote{In our Universe, the uncertainties
in the extraction of the value of the light quark masses are
such that $y_u^\oplus = 0$ is less than 3 standard deviations
away from the central experimental value~\cite{aneesh}.} Our
results, expressed in Eqs.~\eqref{eq:lightratio}
and~\eqref{eq:finalratio}, are cutoff independent, and differ
from those of~\cite{Donoghue:2009me} regardless of the PDF used
for the Higgs VEV.

\medskip

\mysection{Discussion} The gauge hierarchy problem of
Eq.~\eqref{eq:weak-hierarchy} has led to an enormous effort of
``model-building,'' i.e., of proposing new dynamical symmetries
that might explain the smallness of $M_{\rm weak} / \mpl$. But
the inflationary multiverse and the string landscape raise the
disturbing possibility that such efforts could be as misguided
as Kepler's attempts, in the 16th century, to explain the
Titius-Bode law (an approximate mathematical regularity in the
sizes of the orbits in our solar system) in of terms nested
Platonic solids~\cite{titius-bode}. We now know that the laws
of gravity are compatible with a huge variety of solar systems
and that the only significant constraints on the details of our
own solar system, such as the sizes of orbits, are anthropic.

Proposed anthropic explanations of the observed values of the
parameters of the SM, however, should take into account that
many parameters could scan over the multiverse and that the
variation of one might compensate for the variation of another
in such a way that life could be possible in universes
different from ours. The flavor hierarchy puzzle (which {\it a
priori} has nothing to do with the weak scale) suggests that
the Yukawa couplings of the SM may be realized as VEVs of
dynamical fields~\cite{froggatt-nielsen, strong-dynamics,
split-fermions}. If that is the case, then the Yukawa
couplings, though dimensionless in themselves, may scan on the
multiverse along with the dimensionful
quantities~\cite{friendly-landscape}, and fairly generic
considerations about flavor dynamics lead us to expect that
most hospitable universes would lack an observable weak force.
This presents a serious challenge to the notion that the gauge
hierarchy problem can be solved anthropically.

Unless the probability density function of the SM parameters in
the multiverse turns out to have some quite specific features
(such as a strong bias towards small values of the Higgs
self-coupling~\cite{topmass}), the solution to the gauge
hierarchy problem should be {\it dynamical}, and some new,
detectable interactions ---beyond those of an elementary Higgs
field--- should appear in this Universe at energy scales
currently being probed by the Large Hadron Collider.  The
multiverse does not do away with the need for model-building.

\medskip

\mysection{Acknowledgments} We thank Michael Dine, John
Donoghue, Guido Festuccia, Yuval Grossman, Aneesh Manohar,
Takemichi Okui, Mike Salem, Matt Schwartz, and Kris Sigurdson
for discussions.  AJ also thanks Bob Jaffe for encouragement,
Graeme Smith for advice on the manuscript, and the Aspen Center
for Physics for hospitality while some of this work was being
completed. The work of AJ is supported in part by the US
Department of Energy under contract DE-FG02-97IR41022. GP is
the Shlomo and Michla Tomarin career development chair and
supported by the Israel Science Foundation (grant \#1087/09),
EU-FP7 Marie Curie, IRG fellowship and the Peter \& Patricia
Gruber Award.

\end{document}